\documentstyle[twocolumn,prb,aps,epsfig]{revtex}

\def\VEV#1{\left\langle #1\right\rangle}

\begin{document}
\title{
       Conductivity of quantum-spin chains: A Quantum Monte Carlo approach}

\author{J.V. Alvarez and Claudius Gros 
       } 
\address{Fakult\"at 7, Theoretische Physik,
 University of the Saarland,
66041 Saarbr\"ucken, Germany.}

\maketitle

\begin{abstract}
We discuss zero-frequency transport properties of
various spin-1/2 chains. We show, that a careful
analysis of Quantum Monte-Carlo (QMC) data
on the imaginary axis allows to distinguish 
between intrinsic ballistic and diffusive transport.
We determine the Drude weight, current-relaxation
life-time and the mean-free path for integrable
and a non-integrable quantum-spin chain.
We discuss, in addition, some phenomenological
relations between  various transport-coefficients 
and thermal response functions.
     
\end{abstract}


\vspace*{1cm}
\section{Introduction}\label{sect_intro}

The role of the spin excitations on the 
transport properties
of quasi-one dimensional Mott-insulators 
has been the subject of extensive experimental research 
in the last few years. A recent $^{17}$O
NMR investigation~\cite{Thu01} of Sr$_2$CuO$_3$,
extending an earlier $^{63}$Cu NMR study~\cite{Tak96},
measured a $q=0$ spin-diffusion coefficient (equivalent 
to diffusive magnetization transport) several  
orders of magnitude larger than the value for 
conventional diffusive systems. 
Thermal transport measurements in
Sr$_2$CuO$_3$ and SrCuO$_2$ indicate
at the same time, quasi-ballistic transport 
with a mean-free path of several thousands~\cite{Sol00} 
of $\AA$. 

It is well known from structural considerations~\cite{Ami95} and
from studies of the magnetic excitation spectrum~\cite{Ami95,MOTOYAMA},
that Sr$_2$CuO$_3$ and SrCuO$_2$
can be accurately described by the XXZ chain
\[
H^{(xxz)}\ =\ {\sum}_{i} \left[\frac{J_{xx}}{2}
\left(S_{i}^{+}S_{i+1}^{-}+S_{i}^{-}S_{i+1}^{+}\right)
+ J_{z} S_{i}^{z}S_{i+1}^{z}\right]~.
\]
Evidence for ballistic (or quasi-ballistic)
magnetization transport have been found in
recent exact diagonalization 
studies~\cite{BONCA} of $H^{(xxz)}$ at
high temperatures~\cite{Nar98,Fab97}.  
A connection between integrability,
conservation laws and ballistic transport has been proposed
by Zotos and 
coworkers~\cite{Cas95,Zotos-Prelovsek96,Zotos-Naef-Prelovsek97,Naef98,Nar98}.
If the current-current correlation does not decay
to zero for long times, i.e\  when part of the current operator
is conserved, i.e.\ when a certain (non-zero) projection of the
(here magnetic) current operator commutes with the
hamiltonian, the transport is ballistic even at
finite temperatures. This seems to be the case, in general,
for Bethe-Ansatz solvable models like $H^{(xxz)}$, although
a formal proof for this connection is still outstanding.

At present it is unclear, whether there exist 
non-integrable models which do exhibit ballistic
transport, none has been found so far. 
Real compounds like Sr$_2$CuO$_3$ and SrCuO$_2$ correspond
to $H^{(xxz)}$ anyhow only in first approximation.
It is therefore important to examine whether
general, non-integrable, quantum spin chains
show ballistic or diffusive transport properties
at finite temperatures. This question has
been studied by Rosch and Andrei~\cite{ROSCH} 
within a short-time approximation (a memory-matrix approach,
extending an earlier analysis by Giamarchi \cite{Gia92})
for Luttinger-liquids with higher-order Umklapp-scattering.
They found only exponentially small deviations
from ballistic transport away from
commensurability. An alternative route to diffusive 
transport, the coupling spin-phonon coupling  
has been studied  by Narozhny~\cite{Nar96}. 

In his seminal paper~\cite{KOHN} in 1960, W. Kohn proved 
that the existence of a the  delta peak at zero frequency 
(the Drude peak) in the conductivity
is the essential difference between ground states with
localized and extended electronic states.
A simple extension of this idea to the spin transport 
can be used to distinguish, without other explicit
information about the excitation spectrum,
a spin insulator, like spin-Peierls compounds,
from a spin conductor like Sr$_2$CuO$_3$. 
Despite an on-going effort~\cite{SCALAPINO,HANKE} 
devoted to this problem,
the fundamental difference between models with ballistic
and diffusive transport properties
has shown up only recently in QMC-simulations \cite{Alv02}.
The purpose of this paper is to explain in detail how
this important issue can be tackled numerically by QMC.

The organization of the paper is as follows. 
Section \ref{sect_spin_conduct} contains 
the basic definitions and sets the notation that we will 
use along the paper. 
In Section \ref{sect_relation} we deduce the connection between 
the spin current-current and density-density correlation functions
emphasizing the role of boundary terms that occur in Matsubara 
formalism and in Section \ref{sect_QMC}
we discuss how to exploit that connection 
to compute the conductivity in imaginary frequency using 
Quantum Monte Carlo cluster algorithms in an efficient way.
In Section \ref{sect_data_analysis}
we describe a procedure to extract transport coefficients 
(Drude weight and the diffusion coefficient) from QMC-data 
in general  1D interacting systems either 
integrable or non-integrable. In 
Section \ref{sect_ballistic} we apply this method to the XXZ
chain and we obtain the Drude weight at finite temperatures.
We then discuss several phenomenology relations between
transport and thermal coefficients in Section \ref{sect_phenomenology}.
Section \ref{sect_diffusive}
is devoted to the computation of diffusion constants, 
mean-free paths and life-times in a   
non-integrable spin chain. In Section \ref{sect_conclusions} 
we present our conclusions.

\section{Spin conductivity}\label{sect_spin_conduct}

QMC-simulations yield in general correlations functions
on the imaginary-time axis. We therefore consider
the Kubo formula for spin conductivity  in the Matsubara formalism. 

The spin conductivity in one-dimensional spin chains
can be defined as the response of the current to a homogeneous and 
time dependent twist $\Phi=\sum_l\phi_l$ in the quantization axis.   

\begin{equation}
H^{(xxz)}(\Phi) =\ {\sum}_{l}
\big[ k_{l}cos(\phi_l)+j^{z}_{l}sin(\phi_l)
+ J_{z} S_{l}^{z}S_{l+1}^{z}\big]
\label{H_phi}
\end{equation}

where 
\begin{equation}
k_l\ =\ \frac{J_{xx}}{2}\big(S_{l}^{+}S_{l+1}^{-}
                            +S_{l}^{-}S_{l+1}^{+}\big)
\end{equation}
and 
\begin{equation}
j^{z}_l\ =\ \frac{J_{xx}}{2i}\big(S_{l}^{+}S_{l+1}^{-}
                                 -S_{l}^{-}S_{l+1}^{+}\big)
\label{Paramagnetic_current}
\end{equation}
Formally \ (\ref{H_phi}) is the hamiltonian of a XXZ chain in which the 
quantization axis of the  local spin operators 
has been rotated by a site-dependent angle $\phi_{l}$ 
along the z-axis.
To obtain the expression of the Kubo formula for the 
spin conductivity  we expand $H^{(xxz)}(\Phi)$ in a Taylor
series: 
\begin{equation}
H^{(xxz)}(\Phi)\ =\ H^{(xxz)}+{\sum}_{l} j^{z}_l\phi_{l}+\frac{k_{l}}{2} \phi^{2}_{l}
\end{equation}
and we obtain the  {\em total} spin-density current by differentiating 
with respect to $\phi_{l}$.

\begin{equation}
{\partial H^{(xxz)}(\Phi) \over \partial \phi_{l}}\ =\
(j^{z}_{l})^{T}\ =\ j^{z}_{l}+k_{l}\phi_{l}
\label{Total_current}
\end{equation}

The first term is the paramagnetic part of the current.
If the z-component of the magnetization is conserved, 
it can also be deduced using the discretized continuity equation.
\begin{equation}
{\partial\over \partial t} S_l^z(t)\,+\, 
\big(j_{l}^z(t)-j_{l-1}^z(t)\big)\  =\ 0
\label{continuity}
\end{equation}
where the second term is the discrete version of the 
divergence in one dimension. If we combine it 
with the equation of motion
\begin{equation}
{\partial\over \partial t} S_l^z(t)\ =\ i[H,S_l^z]
\end{equation}
we obtain the  expression (\ref{Paramagnetic_current}). The 
second term in Eq.\ (\ref{Total_current}), proportional to the magnetic flux
is called the diamagnetic current. The expectation value of
the total current is
\begin{equation}
\VEV{j^{T}(q,\omega_{n}}\ =\ -\big(\VEV{K} \,+\, \Lambda(q,\omega_{n})\big)\,
\phi_{l}~,
\label{j_T}
\end{equation}
where $\VEV{K}=\VEV{\sum_{l}k_{l}}$ is  the expectation value of the
kinetic energy per site and 
$\Lambda$ is the current-current correlation as a 
function of the Matsubara frequency,
\begin{equation}
  \Lambda(q,\omega_{n})\,=\, 
  \frac{1}{L} \int_{0}^{\beta} e^{i \omega_{n} \tau}
 \VEV{j^{z}(q,\tau)j^{z}(-q,0)}\,d\tau~.
 \label{Lambda}
\end{equation}
The response to the time-integrated twist
is then obtained from (\ref{j_T}) and 
the dynamical conductivity takes the usual form
\begin{equation}
\sigma(q,\omega_n)\ =\ 
{-\VEV{K} \,-\, \Lambda(q,\omega_{n})\over \omega_n}
\ \equiv\ {D(q,\omega_n)\over \omega_n}~.
\label{sigma}
\end{equation}
Eq.\ (\ref{sigma}) leads via analytical continuation
$i\omega_{n} \to \omega+i\delta$ and $\delta \to 0$,
using $\sigma(\omega_n)=\lim_{q\to0}\sigma(q,\omega_n)$,
to the usual representation of the dynamical conductivity
\begin{equation}
\sigma(\omega)\ =\ \pi D(T)\delta(\omega) +\sigma_{reg}(\omega,T)~,
\label{sigma_q_omega} 
\end{equation}
where $D(T)$ is the Drude weight\cite{note_def} 
that can be computed via
$D(T) =\lim_{\omega_{n\rightarrow 0}}\lim_{q\rightarrow0}D(q,\omega_{n})$:

\begin{equation}
D(T) \ = \
 -\VEV{K} \,-\, \Lambda(q \rightarrow 0,\omega_{n}\rightarrow 0 )~.
\label{Drude}
\end{equation}
It can be proved \cite{Zotos-Naef-Prelovsek97} that
this limit is consistent with the definition introduced 
by W. Kohn \cite{KOHN} at $T=0$, 
\begin{equation}
D\ =\ L\left(\frac{d^2 E}{d^2 \Phi}\right)_{\Phi=0}~,
\end{equation}
where $E$ is the ground state energy  and $\Phi$ the total
external flux. Recently $D$ has been extended by 
Zotos, Naef and Prevlosek \cite{Zotos-Naef-Prelovsek97} to finite $T$
\begin{equation}
D(T)\ =\ L\, \sum_{\alpha}\,
{ \exp(-\beta E_{\alpha}) \over Z}\,\left(\frac{d^2 E_{\alpha}}{d^2 \Phi}
                                    \right)_{\Phi=0}
\end{equation}

It is important to note that the limits 
$\lim_{q\rightarrow0}$ and $\lim_{\omega\rightarrow 0}$
do not commute. When the limits are taken in the opposite order 
one obtains the conventional spin stiffness which represents the 
response to a {\em static} twist.  
\begin{equation}
\rho_{s}\ =\ \lim_{q\rightarrow0}\lim_{\omega_{n}\rightarrow 0} D(q,\omega_{n}) 
\end{equation}

A non-zero value of the Drude weight implies that the 
total magnetic-current does not decay to zero when $t \to \infty$  
( i.e. the transport is ballistic). The most simple example
to illustrate this situation is the XX chain. In that case 
$[H_{XX},j^{z}(q)]=0$. Taking the spectral representation
of $\Lambda(q,\omega_{n})$ for $\omega_{n}>0$ we have

\begin{equation} 
\Lambda(q,\omega_{n})
 \ = \  
\frac{1}{ZL}\sum_{m,n}^{E_{m} \neq E_{n}} 
\frac{e^{-\beta E_{m}}| \langle m|j^{z}(q)|n \rangle|^{2}}
      {i\omega_{n}-(E_{m}-E_{n})} 
\label{Spectral_Lambda}
\end{equation}

Note that degenerated states are explicitly  excluded from the sum 
and therefore $|\langle m|j^{z}(q)|n \rangle|=0$ 
and the conductivity reduces to
$\sigma=-\pi\VEV{K} \delta(\omega)$ that saturates the f-sum rule. 
Interactions spoil the 
commutation of spin-current and Hamiltonian but, if the Umklapp
part of the interaction is irrelevant and the system 
remain gapless, the Drude weight remains finite and 
the current-current 
correlation functions can reduce the Drude peak from the kinetic
energy. This situation is indeed realized 
in the gapless regime of the XXZ chain
but the integrable nature of the interaction in this case plays, 
as we will see, a definitive role.  
The regular part of the conductivity 
is in any case ( integrable or non-integrable systems) 
enhanced to fulfill the f-sum rule.
  
\begin{figure}[t]
\noindent
\\
\centerline{
\epsfig{file=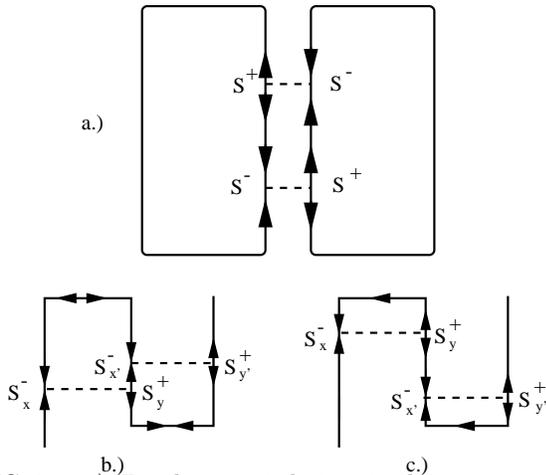,width=0.40\textwidth} 
\vspace{6pt}
           }
\caption{\label{loops} a.) Two-loop contribution to the 
current-current correlation function, a $S^{+}$ and $S^{-}$
operator must be applied in each loop to close it consistently
in terms of the loop orientation. b.) and c.) Two kinds of 
one loop contributions to the current-current correlation function
The ordering of the $S^{+}$ and $S^{-}$ in terms of loop time 
is crucial to evaluate the contribution of these terms.   
}
           
\end{figure}
 
\section {Relation between correlation functions at $T \neq 0$}
          \label{sect_relation}

Now we will derive a connection in between the
current-current 
correlation function $D(q,\omega_n)$ and the dynamical 
susceptibility $S(q,\omega_n)$; 
In the next Section we will explain 
how to exploit this connection in QMC calculations.
The spin-spin correlation function is defined by
\begin{equation}
S(q,\omega_{n})\ =\ {1\over L}\int_{0}^{\beta}
e^{i \omega_{n} \tau}\langle S_{q}^{z}(\tau)S_{-q}^{z}(0) \rangle~.
\label{S_q_o}
\end{equation}
In Fourier space, the continuity equation (\ref{continuity})
takes the form
\begin{equation}
{d\over d\tau}S_q^z(\tau) \ =\
[H,S_q^z] \ =\ i\left( 1-e^{iq}\right) j_q^z~.
\label{cont_q}
\end{equation}  
We integrate the right-hand side of Eq.\ (\ref{S_q_o})
with respect to $\tau$ twice, use (\ref{cont_q}) and obtain
\begin{equation}
S(q,\omega_{n})\ =\ {-1\over\omega_n^2}
\langle\, [[H,S_q^z],S_{-q}^z]\, \rangle
\,-\,{4\sin^2(q/2)\over\omega_n^2} \Lambda(q,\omega_n)~,
\label{S_q_new}
\end{equation} 
where we have used the definition (\ref{Lambda}). 
The double commutator in the right hand side of 
Eq. (\ref{S_q_new}) is the boundary
term of the partial integration and is
evaluated to
\begin{equation}
\langle\, [[H,S_q^z],S_{-q}^z]\, \rangle \ =\ 4\sin^2(q/2)\VEV{K}
\end{equation} 
Recalling the definition of $D(q,\omega_n)$ we arrive  to  
\begin{equation}
%
D(q,\omega_{n})\ =\ 
{\omega_n^2 \over 4\sin^2(q/2)}\, S(q,\omega_n)~.
\label{relation}
\end{equation}
Note, that the double commutator in (\ref{S_q_new})
occurs for the Matsubara
correlation functions and does not occur for a
related real-frequency correlation function \cite{Naef98}.


\begin{figure}[t]
\noindent
\\
\centerline{
\epsfig{file=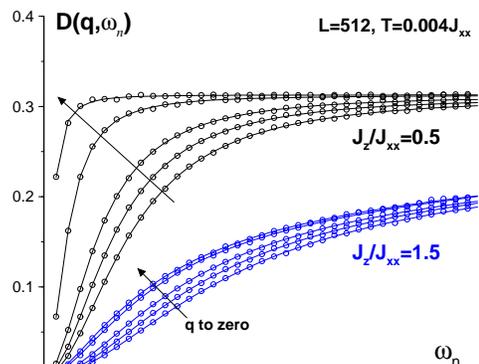,width=0.35\textwidth} 
}
\caption{\label{D_omega}
$D(q,\omega_n)$ as a function of $\omega_n$
for various momenta $q$, $T=0.004J_{xx}$ and
$J_z/J_{xx}=0.5,\ 1.5$. The lines are fits by
Eq.\ (\protect\ref{fitting}).
Statistical errors are of the order of the symbol size.
Note the different limiting behavior of
$D(q\to0,\omega_n)$ for $J_z/J_{xx}=0.5$
and $J_z/J_{xx}=1.5$.
}
           
\end{figure}
\section{QMC evaluation of the conductivity}\label{sect_QMC}

In this Section we will discuss the usefulness of Eq.\ (\ref{relation}) in
the context of QMC-simulation, comparing
two different possibilities to compute the conductivity
using quantum cluster algorithms. 

Cluster algorithms for QMC-simulations
allow for global updates of the configuration 
by flipping simultaneously spin-clusters whose
typical sizes are of the order of the
correlation length of the system. 
The loop algorithm\cite{EVERTZ} we used in the present study
gives an efficient prescription to construct clusters.
The resulting autocorrelation time is in general
of the order of one Monte Carlo step (see \cite{EVERTZ_REV}
for an excellent review). 

The current-current correlation function in real space 
and imaginary time takes the form
\begin{equation}
\Lambda(l,\tau)\ = \ {1 \over L N_{T}}
\sum_{l',\tau'}\, j_{l+l'}^{z}(\tau'+\tau)\,j_{l'}^{z}(\tau')
\label{Lambda_real_space}
\end{equation}

where $N_T$ is the number of Trotter slices.
The contributions to $\Lambda(l,\tau)$ are non-diagonal four-site
operators, typically 
$(J_{xx}/4) S^{+}_{l_1}(\tau_1)S^{-}_{l_1+1}(\tau_1)
S^{-}_{l_2}(\tau_2)S^{+}_{l_2+1}(\tau_2)$.
In principle non-diagonal operators can be computed using 
the loop algorithm\cite{WIESE}. When these non-diagonal operators
are two-point-like only one-loop terms contribute 
to the correlation function. In that case it is possible 
to design efficient improved estimators, meaning that a given magnitude 
is evaluated not only in one configuration but in all configurations 
related by loop flippings.
The evaluation of a four-point correlation function is more involved
\cite{ALVAREZ00}. In that case there are two-loop terms
and one-loop terms which contribute in different ways depending
on the specific shape of the loop, see Fig.\ \ref{loops} for an
illustration.
As a consequence the improved estimators are much less efficient.
The dynamical susceptibility in  $S(q,\omega_{n})$
is, on the other hand, a two-point diagonal operator 
that can be evaluated efficiently using improved estimators
and it is related to the conductivity
using the Kubo formula (\ref{sigma}) and the relation (\ref{relation}).

In particular one can compute within each loop $\alpha$ the magnitude: 
\begin{equation}
W(q,\omega_{n},\alpha)\ =\ { \beta \over N_{T}} \sum_{(x,\tau)\in\alpha} 
S^{z}_{l}(\tau)\,e^{i(ql+\omega_{n}\tau)} ~.
\end{equation}
The dynamical structure factor is then
\begin{equation}
S(q,\omega_n)\ \sim\ \sum_{\alpha}\, W(q,\omega_{n},\alpha)
                                     W(-q,-\omega_{n},\alpha)~,
\end{equation}
where $\alpha$ runs over all loops constructed.
In particular we want to emphasize the importance of relation  
\ (\ref{relation}), because only using it we obtained the
high-quality data (large set of uncorrelated measurements 
with  small statistical error bars)  that is necessary
in order to extract DC-transport coefficients.

A second technical important issue is the 
relation between the conductivities in real and imaginary axis.
As has been discussed  by Kirchner, Evertz and Hanke the limit 
$\omega \to 0$ of the conductivity can be also approached from 
the imaginary frequency axis \cite{HANKE}. 
Taking the analytical continuation to the real frequencies 
$i\omega_{n} \to \omega+i\delta$ in the spectral representation
of $\Lambda(q,\omega_{n})$ \ (Eq. \ref{Spectral_Lambda})
we note  that $\Lambda(q,\omega+i\delta)$ is analytic in the
upper half of the complex $\omega$-plane. Zero-frequency
properties like the Drude weight or the diffusion constant can be 
reliably extracted by the extrapolation
along the imaginary axis at low
temperatures, when many Matsubara frequencies
$\omega_n=2\pi T n$ are available close to $\omega=0$ for the
extrapolation. This is the case, however, only at
low temperatures and will therefore present results
only for $T\ll J_{xx}$.
     
Finally we mention a few numerical details.
We used the discrete imaginary-time version of the Loop Algorithm 
with a Trotter decomposition of typically  
$N_T=800-2000$ and on the average
$6\times 10^6$ full MC-updates in the grand-canonical
ensemble. Test-runs within a canonical ensemble
were also performed to exclude any influence 
of the ensemble in the transport properties. 
The error bars (either the statistical ones  or those derived from the fitting)
are of the order of the symbol size in all the figures presented.


\begin{figure}[t]
\noindent
\\
\centerline{
\epsfig{file=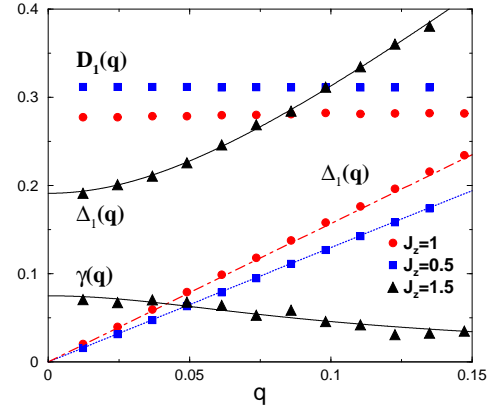,width=0.35\textwidth} 
}
\caption{\label{par_Jz_q}
$D_1(q)$, $\Delta_1(q)$ and $\gamma_1(q)$ from Eq.\ (\ref{fitting})
as a function of momenta $q$ for the XXZ-model, $L=512$ and 
for various $J_{z}$ at $T=0.004\,J_{xx}$.  $\gamma_1(q)$ is too small 
for $J_z\le J_{xx}$ to show up on this scale.
The lines are the Bethe-Ansatz result (\ref{c_exact})
for the velocity $c(J_z)$ (no fit, for $J_z\le J_{xx}$).
For the discussion of the fit for $J_z=1.5 J_{xx}$ see
the text.
}
           
\end{figure}

\section {Data-Analysis}\label{sect_data_analysis}

At low temperatures and frequencies, the scaling of $D(q,\omega_{n})$
can be obtained simply invoking the conformal symmetry 
of the model emerging  in the gapless regime $J_{z}<J_{xx}$.
$S(q,\omega_{n})$ at small q takes then form
\begin{equation}
S(q,\omega_{n})\ =\  \frac{D_1(T)q^{2}}{(cq)^2+\omega_{n}^{2}}.
\label{LL_S_q}
\end{equation}
Note that, unlike near $q=\pi$, the dynamical susceptibility
about $q=0$ do not show power laws. 
The XXZ-model maps to an interacting 1D spinless fermionic 
system at half filling. For the noninteracting case 
(the XX chain) we can compute exactly $D_1(q,\omega_{n})$
and we obtain $D_1(0)=J_{xx}/\pi$, 
and  $\lim_{q\to0}J_{xx}\sin(q)= J_{xx} q\equiv cq$. 

Expression (\ref{LL_S_q})
and Eq.\ (\ref{relation}) suggest the form
\begin{equation}  
 D(q,\omega_{n})\,=\,
 \frac{D_1(T)\,\omega_{n}^{2}}{\Delta^{2}(q)+\omega_{n}^{2}}~.
\label{LL}
\end{equation}
Alternatively, Eq.~(\ref{LL}) can be viewed as the first term
of the exact representation for $D(q,\omega_{n})$ containing
an infinite-number of terms \cite{HANKE}:
\begin{equation}     
D(q,\omega_{n})\ =\ \sum_{j=1}^{2}
\frac{D_j(q)\,\omega_{n}^{2}}
{\Delta_j^{2}(q)+2\gamma_j(q)\,\omega_n+ \omega_{n}^{2}}~.
 \label{fitting} 
\end{equation}    
The choice of this fitting function is essential 
to distinguish the transport properties of 
ballistic and diffusive systems, indeed 
it allows the correct computation of the Drude weight and  
the diffusion coefficient. We discuss now in detail the
properties of (\ref{fitting}):
     
\noindent
{ \bf i.) } $D(q,\omega_{n})$ is analytic in the upper complex-plane
     for $\gamma_j(q)\ge0$.

\noindent
{ \bf ii.) } For the zero-$q$ gaps
             $\Delta_i(0)=\lim_{q\to0}\Delta_i(q)$ we
	     find to possibilities:
             (a) $\ \Delta_1(0)=0\ $ and $\ \Delta_2(0)>0$,
	     i.e.\ Eq.\ (\ref{fitting}) describes a gapless phase.
	     (b) $\ \Delta_1(0)>0\ $ and $\ \Delta_2(0)>\Delta_1(0) $ 
	     and i.e.\ Eq.\ (\ref{fitting}) describes a gaped phase.
    In the first case, (\ref{fitting}) reproduces 
    the correct $\omega$ and $q$ dependence 
    for the scaling form of a the Luttinger liquid (\ref{LL}).
    The first term in Eq.\ (\ref{fitting}) dominates the
low-frequency behavior in both cases and we have set
generally $\gamma_2\equiv0$ in order to keep the
number of parameters to a minimum.

\noindent
{ \bf iii.) }  At high frequencies  
\begin{equation}     
\lim_{\omega_n\to\infty}D(0,\omega_{n})\ =\  -\VEV{K}\ \equiv\
D_1(0)+ D_2(0)
\label{limit_large_omega_n}
\end{equation}
and a finite $D_2(0)$ results in a reduction of 
the Drude weight $D(T)$ with respect to the kinetic energy,
see Eq.\ (\ref{Drude}).
A finite $D_2(0)$ measures therefore the amount of
decay experienced by the {\em total} current due to the interactions.
We note also  that the Ansatz
Eq.\ (\ref{fitting}) for $D(q,\omega_n)$, 
together with Eq.\ (\ref{limit_large_omega_n}), 
is consistent with the f-sum-rule \cite{Bae79}
\begin{equation}
{1 \over\pi}\int_0^\infty \mbox{Re}\,\sigma(\omega)\,d\omega\ =\ -\VEV{K}
\label{f-sum-rule}
\end{equation}
for the optical conductivity.

\noindent
{ \bf iv.) }  In the  gapless regime $D(q,\omega_{n})$ can describe a normal
conductor with finite DC-conductivity. 
The optical conductivity (\ref{sigma}) takes for small 
frequencies the Drude form.
\begin{equation}
\mbox{Re}\,\sigma(\omega)\ =\ 
{2D_1(0)\gamma_1(0)\over \omega^2+4\gamma_1^2(0)}
\ \equiv\
{\sigma_0\over 1 + (\omega\tau)^2}~,
\label{Drude_formula}
\end{equation}
where we introduced the
DC-conductivity
\begin{equation}
\sigma_0\ =\ D_1(0)/(2\gamma_1(0))
\label{DC-conductivity}
\end{equation}

and the quasi-particle lifetime
\begin{equation}
\tau\ =\ (2\gamma_1(0))^{-1}~.
\label{life-time}
\end{equation}

For $\tau\to\infty$
Eq.\ (\ref{Drude_formula}) reduces to
$\mbox{Re}\,\sigma(\omega)= \pi D_1(0)\,\delta(\omega)$.
Even more, if we consider now  the q-dependence
for $1/\omega\gg \tau$, the optical 
conductivity takes (for small $cq/\gamma_1(0)$) the 
diffusion form
\begin{equation}
\sigma(q,\omega)\ =\ 
{\sigma_{0}\,\omega\over \omega +i D_s\,q^2}~,\quad
D_s = {c^2\over2\gamma_1(0)}\equiv c^2\tau~.
\label{diffusion}
\end{equation}
$D_s$ is the spin-diffusion constant. Eq.\ (\ref{diffusion})
is consistent with $D_s=c\lambda_s$, where
$\lambda_s=c\tau$ is the mean free length. 

\noindent
{ \bf v.) } The uniform spin stiffness 
$\rho_{s}\ =\ 
\lim_{q\rightarrow0}\lim_{\omega_{n}\rightarrow 0} D(q,\omega_{n}) $
is always zero, as expected for a quantum-critical antiferromagnetic chain.   

\noindent
{ \bf vi.) } The quality of the fit to $D(q,\omega_n)$ by
Eq.\ (\ref{fitting}) is, in general, excellent, as illustrated
in Fig.\ \ref{D_omega} for $J_z=0.5J_{xx}$ and $J_z=1.5J_{xx}$.
Note that the $q\to0$ limiting curve for $D(q,\omega_n)$ is
singular in the gapless phase ($J_z=0.5J_{xx}$), but well defined
in the gaped phase (($J_z=1.5J_{xx}$).


\begin{figure}[t]
\centerline{
\epsfig{file=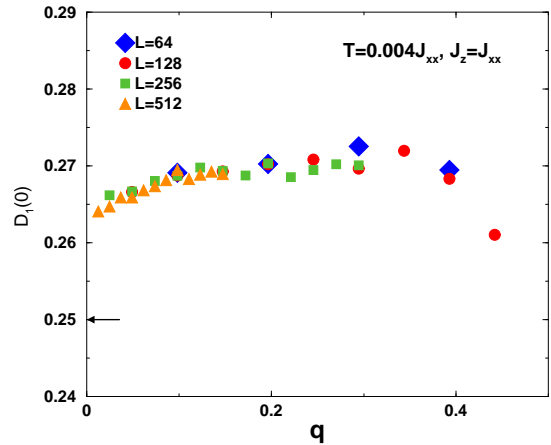,width=0.40\textwidth} 
           }
\vspace{4pt}
\caption{\label{xxx_D1_L}
The q-dependent Drude weight $D_1(q)$ for
the isotropic Heisenberg chain at $T=0.004J_{xx}$
for various system sizes $L=64,\dots512$.
The $T=0,\ q=0$ result given by Eq.\ (\protect\ref{D(0)})
is indicated by the arrow. The convergence with system
size is slow for $q\to0$, due to the logarithmic corrections
present at the isotropic point $J_z=J_{xx}$.
}
\end{figure}


\section {Ballistic transport}\label{sect_ballistic}
 
In this Section we will apply the procedure 
described in Section \ref{sect_QMC} to the XXZ chain. We will
compare with exact known results and study the controversial 
finite temperature behavior of $\sigma(\omega=0)$ 
for this model.
 
In Fig.\ \ref{par_Jz_q} we show the values 
for $D_1(q)$ and $\Delta_1(q)$ and $\gamma_1(q)$
for $J_z/J_{xx}=0.5,\ 1.0$ and $1.5$. These are the
values used in Fig.\ \ref{D_omega} for fitting
$D(q,\omega_n)$.
We have included (no fit) in Fig.\ \ref{par_Jz_q},
for the gapless regime $J_z\le J_{xx}$, the
Bethe-Ansatz result 
$\lim_{q\to0}\Delta_1(q)= c(J_z)\,q$
for the magnon dispersion,
where $c(J_{z})$ is the velocity 
\begin{equation}
c(J_z) \ =\ {\pi\over2}\,
{\sqrt{J_{xx}^2-J_z^2}\over 
\mbox{arcos}(J_{z}/J_{xx})} \ =\ 
{\pi\over2}\,{\sin(\theta)\over\theta}
\label{c_exact}
\end{equation}
of the des Cloiseaux-Pearson spectrum \cite{Cloi66},
with $J_z=\cos(\theta)J_{xx}$. In the gaped phase
we have fitted $\Delta_1(q)$ 
by $\varepsilon(q)=\sqrt{\Delta_0^2+(cq)^{2}}$. We find
$\Delta_0=0.191J_{xx}$, which is close to twice the 
one-magnon gap of $0.091J_{xx}$ \cite{Cloi66}.

The damping $\gamma_1(q)$ is vanishing small
for $J_{z}<J_{xx}$ and acquires a finite value in
the gaped phase. We found phenomenological that 
$\ \gamma_1(q)\Delta_1(q)\approx\,const.$, 
independent of $q$, for $J_z>J_{xx}$.
In Fig.\ \ref{xxx_D1_L} we present values obtained by QMC
for the $q$-dependent Drude weight for
$J_z=J_{xx}$ at $T=0.004J_{xx}$. We find good
convergence for small but finite-$q$, but slow convergence
for $q\to0$ as a function of system size, due to the
multiplicative logarithmic corrections present at the isotropic point.
We have indicated by the horizontal arrow the
$T=0,\ q=0$ Bethe-Ansatz result~\cite{SHASTRY}.
For $J_z<J_{xx}$ the agreement in between
low-$T$ QMC and the $T=0,\ q=0$ Bethe-Ansatz result
is excellent~\cite{Alv02}.

\begin{figure}[t]
\centerline{
\epsfig{file=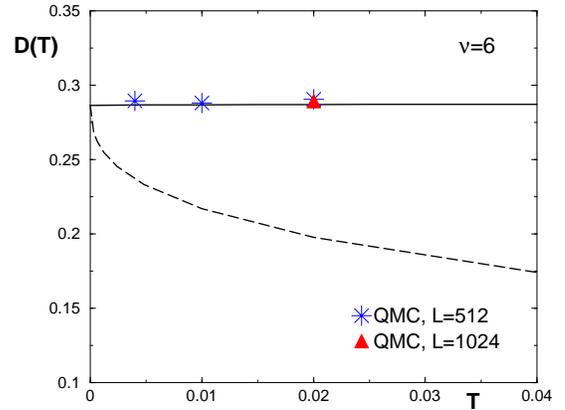,width=0.40\textwidth} 
           }
\vspace{4pt}
\caption{\label{D_T}
QMC results for the Drude weight for 
$L=512,1024$ and $J_z=J_{xx}\cos(\pi/6)=0.866J_{xx}$
as a function of temperature (in units of $J_{xx}$)
in comparison with two (solid lines: Ref.\ \cite{KLUEMPER},
dashed lines: Ref.\ \cite{ZOTOS99}) Bethe-Ansatz results.
}
\end{figure}

\begin{figure}[t]
\centerline{
\epsfig{file=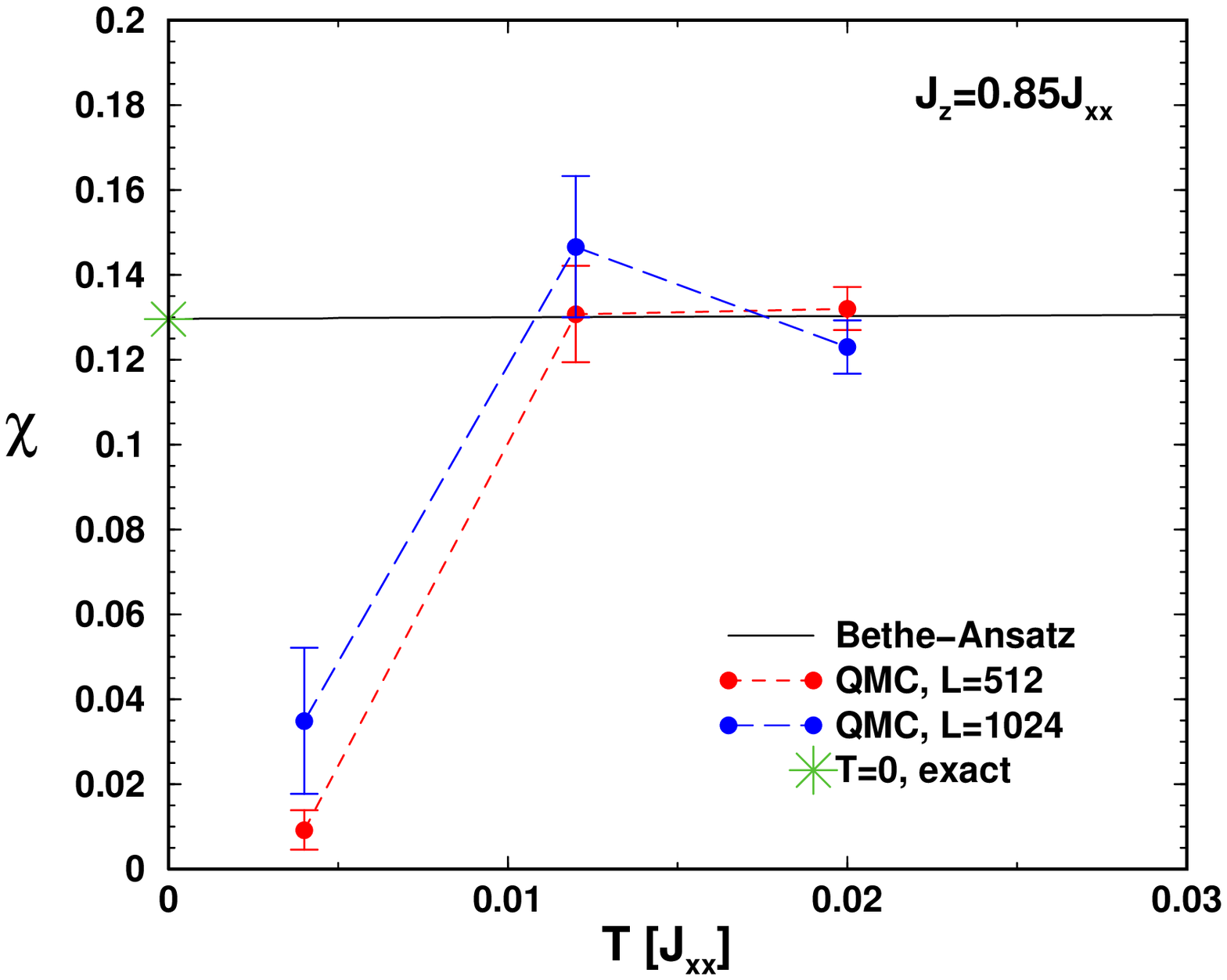,width=0.40\textwidth} 
           }
\vspace{4pt}
\caption{\label{test_chi}
QMC results for the uniform suszeptibility $\chi(T)$
for $L=512,\ 1024$ and $J_z=0.85J_{xx}$ together with the
Bethe-Ansatz result (solid line, Ref.\ \cite{Fab99}),
as a function of temperature.
The star denotes the $T=0$ Bethe-Ansatz result.
Note the absence of finite-size effects for
$T>0.012J_{xx}$ in the QMC data.
}
\end{figure}

We study now the behavior of the Drude weight at finite
temperatures for models free from strong 
multiplicative corrections. 
The main conclusion of a Bethe-Ansatz 
calculation by Zotos~\cite{ZOTOS99} 
is a fast decay of the Drude weight when 
the temperature increases, in agreement with
exact diagonalization studies in the limit
of infinite temperatures~\cite{Nar98}.
Kl\"umper {\it et al.} have found~\cite{KLUEMPER}, 
with an alternative Bethe-ansatz approach, a functionally different
behavior for $D(T)$, see Fig.\ \ref{D_T}.
For a numerical probe of $D(T)$ we focus
on $J_z=J_{xx}\cos(\pi/6)$ and consider several small temperatures.  
For this value of $J_z/J_{xx}$ the numerical problems due
to multiplicative logarithmic corrections are absent
(compare Fig.\ \ref{test_chi})
and the difference in between the two different
Bethe-Ansatz predictions are substantial.
In Fig.\ (\ref{D_T}) we show a comparison of our data with
the two available analytical results \cite{ZOTOS99,KLUEMPER}.
Our results agree with the temperature-dependence predicted
by Kl\"umper {\it et al.}.

We have evaluated the uniform susceptibility $\chi(T)$ for
$J_z=0.85J_{xx}$, $L=512,\ 1024$ and several low temperatures
in order to address the two questions: (a) Is it correct
to compare $T=0$ Bethe-Ansatz results with
QMC-results obtained for a temperature $T=0.004J_{xx}$
and $L=512$? (b) Is $T=0.02J_{xx}$ large enough
for $L=512,\ 1024$ not to be affected substantially
by finite-size effects? The data presented in
Fig.\ \ref{test_chi} shows that $T=0.004J_{xx}$ is indeed
below the finite-size gap and should be a good approximation
to the $T=0$ data and that for $T\ge0.012J_{xx}$
no finite-lattice effect can be observed within the
statistical error-bars given. Note that the
low-T dependence of the Bethe-Ansatz~\cite{Fab99} 
result shown
in Fig.\ \ref{test_chi} for
$\chi(T)$ can be fitted by $\chi(T)\sim T^x$
with $x\approx0.867$. This exponent is very close
to the exact value $x=0.858$ obtained by
Eggert {\it et al.}\cite{EGGERT} for $J_z=0.85J_{xx}$.
Since $x<1$ the slope ${d\chi(T)\over dT}$ is diverging
for $T\to0$. This divergence is, however, not relevant
for the temperature-scale presented in Fig.\ \ref{test_chi}.



\section{phenomenological relations in between transport coefficients}
         \label{sect_phenomenology}

Let us consider a 1D system system with a finite magnetization
relaxation time $\tau$, which might be due to either 
intrinsic relaxation processes or due to weak residual
coupling to an external bath (i.e.\ phonons). The
DC-magnetic conductivity takes then the form
\begin{equation}
\sigma(\omega=0)\ =\  D_1(0)\,\tau~,
\label{sigma_tau}
\end{equation} 
see Eq.\ (\ref{DC-conductivity}) and (\ref{life-time}).

Let us assume that an inhomogeneous magnetic field $B^z(x)$
is applied in the chain along the z-axis. In principle this 
is a non-equilibrium situation but if $d B^z/d x$ is small    
we can assume, after coarse-graining the chain, 
that we have a well defined local magnetization  
$M(x)$ and all thermodynamic relations hold locally. 
Generalizing the usual phenomenology~\cite{ASHCROFT}
for electric transport we write the magnetic current
as
\begin{equation}
j(x)\ =\ v \big( M(x+\lambda)- M(x)\big)\ =\ \lambda v\, \frac{d M}{d x}~,
\end{equation}
where $\lambda=v\tau$ is the mean-free path $v$ is the velocity 
associated to  the magnetization current that we will identify latter.
We can express the magnetization current in terms of the gradient 
of the magnetic field:
\begin{equation}
j(x)\ =\ v^2 \tau\, \frac{d M}{d x}\ =\ v^2 \tau\, \frac{d M}{d B^z}\frac{d B^z}{d x}~.
\end{equation}
Using $j(x)=\sigma {dB^z\over dx}$ we
arrive at  the expression
\begin{equation}
\sigma\ =\ v^2 \tau\, \chi~.
\label{phenomenological}
\end{equation}
This relation is analogous to the well-known~\cite{ASHCROFT}
phenomenological kinetic formula for the thermal conductivity
$\kappa=c_{V} v^2 \tau$,  where $c_{V}$ is the specific heat
and $\kappa$ the thermal conductivity. $\tau$ 
can be eliminated if we use Eq.\ (\ref{sigma_tau}):

\begin{equation} 
{\sigma\over \chi\,\tau}\ =\ {D_1(0)\,\tau\over \chi\,\tau}\ =\ 
{D_1(0)\over \chi}\ =\ v^2~.
\label{ratio_D_chi} 
\end{equation}
This phenomenological equation is independent of the value of $\tau$,
and holds also in the limit $\tau\to\infty$, when $D_1(0)$ becomes
the Drude weight.
Since the derivation of (\ref{ratio_D_chi}) is based on quasi-ballistic
arguments, it is of interest to examine whether this relation holds
at low-temperatures for Luttinger liquids and Bethe-solvable models.
At $T=0$ both magnitudes $\chi$ and $D$ have been computed exactly 
\cite{SHASTRY,EGGERT} for the XXZ chain by Bethe-Ansatz:
\begin{equation}
D(0) \ =\ {J_{xx}\pi \over 4}\, {\sin(\theta) \over \theta(\pi-\theta) }~,
\label{D(0)}
\end{equation}
were we have defined $J_z=\cos(\theta)J_{xx}$ and
\begin{equation}
\chi(0)\ =\  {\theta \over J_{xx}\pi(\pi-\theta)\sin(\theta)}~.
\label{chi(0)}
\end{equation} 
dividing both relations we obtain:
\begin{equation}
{D_1(0) \over \chi(0)}\ =\ {\pi^{2}J_{xx}^2 \over 4} 
{\sin^{2}(\theta) \over \theta^{2}} \ \equiv\ c^{2}(J_z)~.
\label{D_exact}
\end{equation}      
This results then allows us
to identify the magnetization-transport
velocity $v$ in (\ref{ratio_D_chi}) with the spin-wave
velocity $c(J_z)$: At $T=0$ Eq.\ (\ref{ratio_D_chi}) 
is then exact.
The validity of (\ref{D_exact}) in leading, low-T 
correction is an open question presently.
The leading $T$-correction to $D_1(T)$ and $\chi(T)$
are $\sim T^2$ for $J_z=0$ and do not cancel;
Eq.\ (\ref{D_exact}) is exact for the XX-model
only at $T=0$. The leading $T$-corrections to
the susceptibility show~\cite{EGGERT}, however, an exponent
crossover for $J_z=0.5J_{xx}$ and (\ref{D_exact})
might hold in leading low-T order for $J_z>0.5J_{xx}$.

A relation similar to (\ref{ratio_D_chi})
has been discussed recently for thermal 
transport~\cite{ALVAREZ02,KLUEMPER01}
experiments~\cite{Sol00,Hes01}, 
where we define~\cite{ALVAREZ02,note_def}
$\kappa(T) \equiv \kappa^{th}(T)\,\tau$, where 
$\kappa^{th}$ is the {\em thermal Drude weight}.
For the XXZ chain one finds~\cite{KLUEMPER01} that both
$\kappa^{th}(T)$ and $c_V(T)$ are linear in temperature
for small temperatures and that
\begin{equation}
\lim_{T\to0}{\kappa^{th}(T) \over c_{V}(T)}\ =\ c^{2}(J_z)~.
\label{ratio_kappa_cV} 
\end{equation}  
Combining Eq.\ (\ref{ratio_D_chi}) and
Eq.\ (\ref{ratio_kappa_cV}) we obtain
\begin{equation}
{D_1(0) \over \chi(0)} \ =\ \lim_{T\to0}{\kappa^{th}(T) \over c_{V}(T)}~.
\label{both_ratios} 
\end{equation}  
For ballistic systems the quantity $D_1(0)$ in above equation
is identical to the Drude weight.  This relation can therefore 
be interpreted in the framework of a Luttinger liquid.
The Hamiltonian of a Luttinger liquid can be written in the 
diagonal form:
\begin{equation}  
H\ =\ \sum_{q} v_{s}|q| b^{\dagger}_{q} b_{q}
   +{1 \over 2}{\pi \over L}(v_{N}N^{2}+v_{J}J^{2})~,
\label{Luttinger_liquid}
\end{equation} 
where the first term corresponds to the bosonic part 
and  N and J are integer quantum numbers associated to states 
with nonzero charge and current respectively. The three velocities 
present in the Hamiltonian, the sound velocity 
$v_{s}$, the charge velocity $v_{N}$ and the current velocity $v_{J}$
are not independent but restricted  by an universal relation valid  in all 
microscopic models in the Luttinger-liquid universality class~\cite{HALDANE80}:
\begin{equation}   v_{N}v_{J}\ =\ v_{s}^{2}~.
\label{velocities}
\end{equation} 

For the XXZ chain the values of these three 
parameters of the effective low-energy Hamiltonian (\ref{Luttinger_liquid}) 
have been  identified {\em independently}, $v_{N}$ and $v_{s}$
by Haldane~\cite{HALDANE80} and $v_{J}$ by Gomez-Santos~\cite{GOMEZ-SANTOS} 
using the results of different Bethe-Ansatz studies 
\cite{YANG-YANG,JOHNSON,SHASTRY}.

Guided by the phenomenological derivation presented above  
we propose the following (phenomenological) finite-temperature extensions 
of the velocities in Eq.\ (\ref{Luttinger_liquid}). 
 
\begin{equation} 
v_{N} \to {1\over\chi(T)}, \quad  {v_{J}\over\pi} \to D_1(T) \quad  
v_{s}^{2} \to {\kappa^{th}(T) \over c_{V}(T)} 
\end{equation} 

The extension to finite temperatures of $v_{N}$ and $v_{J}$  
are in  agreement with their physical meaning
$v_{N}=(\frac{d B}{d M})_{B=0}$ 
and~\cite{note_def,GOMEZ-SANTOS} 
$v_{J}/\pi=L(\frac{d^{2}E}{d^{2}\Phi})_{\Phi=0}$ for
Luttinger liquids.
On the other hand, for the magnitudes involved in the  
thermal ratio, the bosonic part of the Hamiltonian 
does play an important role. In fact, only
the bosonic degrees of freedom
transport the energy in the homogeneous
states (which, by definition~\cite{ASHCROFT} do not
carry particle currents) relevant for
the thermal conductivity $\kappa=\kappa^{th}\,\tau$ 
and the specific heat $c_V$.
It is therefore justified to consider $\kappa^{th}/c_{V}$ 
as the natural extension of $v_{s}^{2}$ to low temperatures.
Recently this ratio has been computed using Bethe-Ansatz 
techniques at all temperatures by Kluemper and Sakai \cite{KLUEMPER01}.
$\kappa^{th}/c_{V}$ is a well behaved function of T, even more
it is very flat at low temperatures and takes the value the 
expected value $v_s^{2}\equiv c^2(J_z)$ at $T=0$.  



\section{Diffusive transport}\label{sect_diffusive}

\begin{figure}[t]
  \centerline{
  \epsfig{file=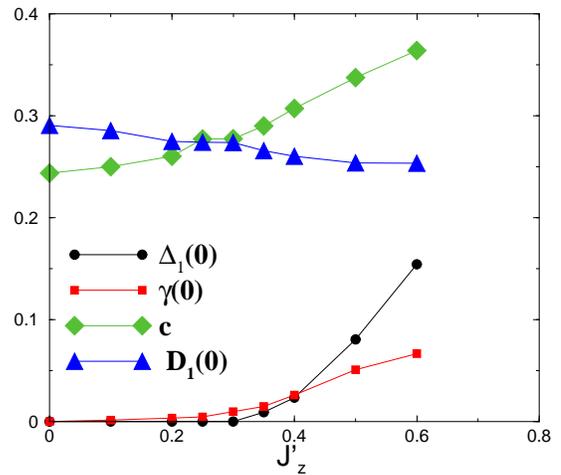,width=0.40\textwidth} 
             }
  \vspace{4pt}
  \caption{\label{par_Jz_prime}
  For $L=512$ and $T=0.004J_{xx}$ the
  QMC-results for the gap $\Delta_1(0)$, the relaxation
  rate $\gamma_1(0)$, the effective velocity $c$ and the parameter 
  $D_{1}(0)$ as a function of $J_{z}'$
  for $H^{(xxz)}+H'$ with $J_z=J_{xx}\cos(\pi/6)$. 
  }
  \end{figure}

We study now the effect of non-integrable interaction 
terms in the conductivity of a 1D spin system.
To be specific, we add a small perturbation
to $H^{(xxz)}$, which breaks the integrability 
of $H^{(xxz)}$:
\begin{equation}
H'\ =\ J'_{z}\,{\sum}_{i}S^{z}_{i}S^{z}_{i+3}~.
\label{Hnonint}
\end{equation}
The expression (\ref{Paramagnetic_current}) for the spin current remains
valid, since $H'$ does contain only $S^z$-operators 
and the system remains non-frustrated and free from sign problems.
We have performed QMC simulations for the resulting model $H=H^{(xxz)}+H'$
mainly for $J_z=J_{xx}\cos(\pi/6)$. We find 
a transition to a gaped phase around
$J_z'\cong 0.3\,J_{xx}$, see Fig.\ \ref{par_Jz_prime}.
The exponential opening of the gap resembles a Kosterlitz-Thouless
transition very similar to the one present  in the XXZ chain 
at the isotropic point and suggests that $H'$ is not changing 
the universal properties of $H^{(xxz)}$, only shifting the 
transition point and adding the ingredient of non-integrability.     
We find the relaxation time 
$\tau=1/(2\gamma_1(0))=\lim_{q\to0}1/(2\gamma_1(q))$ to be finite 
within the numerical accuracy (due to finite-$q$ and $\omega_n$
resolution), leading to a finite DC-conductivity in the
gapless phase.

We have examined the temperature-dependence
of the resulting DC-conductivity. Due to our restriction to
$T\ll J_{xx}$, resulting from the finite-$\omega_n$ resolution
on the imaginary axis (see Section \ref{sect_QMC}) we could 
not examine a large enough $T$-range in order to determine
the full $T$-dependence of $\sigma(0)$. We found
for $J_z' = 0.3 \,J_{xx}$:
 $\sigma(T=0.004J_{xx})=13.6 \pm 0.9$, 
 $\sigma(T=0.008J_{xx})=12.1 \pm 1.0$ and
 $\sigma(T=0.012J_{xx})=10.1 \pm 0.8$.

In agreement with our expectation of a diverging
DC-conductivity in a translational-invariant system
we find $\sigma(0.008)>\sigma(0.012)$. The
increase from $\sigma(0.008)$ to $\sigma(0.004)$ is,
on the other hand, only modest, presumably due to
the finite-size resolution limitation illustrated
in Fig.\ \ref{test_chi}.
 

\section{Conclusions}\label{sect_conclusions}
We have shown that Quantum-Monte-Carlo simulations
of quantum-spin chains are a powerful tool to
obtain finite and diverging transport coefficients
at very low temperatures. We have derived an useful
relation between the dynamical structure
factor $S(q,\omega_n)$ and the dynamical 
conductivity $\sigma(q,\omega_n)$, 
which allows to calculate $\sigma(q,\omega_n)$ to
very high accuracy on the imaginary axis.
For an integrable chain we
support the original suggestion by Zotos {\it et al.}
\cite{Cas95,Zotos-Prelovsek96,Zotos-Naef-Prelovsek97}
of a finite Drude weight at finite temperatures and
settle a recent dispute regarding the functional form
of $D(T)$. In addition we present results suggesting
the absence of ballistic transport (i.e.\ a zero Drude-weight)
for a non-integrable model, for which we are able
to estimate the magnitude of the DC-conductivity.
We have discussed our result in the framelight of
phenomenological relations and Luttinger-liquid theory.
Connections to recent studies of the diverging
thermal conductivity of quantum-spin chains was
made.
  

\section{Acknowledgments}

This work was supported by the DFG. We  
would like to thank A. Kl\"umper and X. Zotos for giving us their
numerical results and their valuable comments. We also 
want to thank R.~Valent{\'\i}
a careful reading and several suggestions.



\begin{thebibliography}{99}
\bibitem{note_def} Note that $D$ and $2\pi D$ in front of
                     the $\delta(\omega)$-function in
                     Eq.\ \protect\ref{sigma_q_omega} are also used
		     in literature and that in Bethe-Ansatz studies the
		     Hamiltonian is defined normally with a prefactor
		     $2J$. With regard to thermal transport,
		     note that $\pi\kappa^{th}\equiv\tilde\kappa$ with
		     respect to the notations used in Ref.\
		     \protect\cite{ALVAREZ02} and Ref.\ 
		     \protect\cite{KLUEMPER01}.
\end{thebibliography}
\end{document}